\documentclass{article}
\usepackage{spconf,amsmath,graphicx,hyperref}

\usepackage[T1]{fontenc}
\usepackage{changes}
\usepackage[noadjust]{cite}
\usepackage{cite}
\usepackage{multicol,lipsum}
\usepackage[shortlabels]{enumitem}
\usepackage{tikz,pgf} 
\usetikzlibrary{arrows,automata} 
\usepackage{verbatim}
\usetikzlibrary{positioning,shapes.geometric}
\usepackage{changes}
\usepackage{amsmath}
\usepackage{algpseudocode}
\usepackage{tabularx}
\usepackage{array}
\usepackage{float}
\usepackage{amsmath}
\usepackage{amssymb}
\usepackage{multirow}
\usepackage{algorithm}
\usepackage{mathtools,xparse}
\DeclarePairedDelimiter{\norm}{\lVert}{\rVert}
\usepackage{amssymb}
\usepackage{gensymb}
\usepackage{array}
\usepackage{tabularx}
\usepackage{enumitem}
\usepackage{lipsum}
\usepackage{color}
\usepackage{graphicx}
\usepackage[caption=false]{subfig}
\usepackage{amsmath}
\usepackage{multirow}
\usepackage[all,poly]{xy}
\usepackage{algpseudocode}
\usepackage{array}
\usepackage[]{siunitx}
\usepackage{textcomp}
\usepackage{nicematrix}
\usepackage[english]{babel}
\usepackage{blindtext}
\usepackage{adjustbox}
\usepackage{cite}
\title{Bayesian Uncertainty-Aware MRI Reconstruction}

\name{Ahmed Karam Eldaly$^{(1),(2)}$, Matteo Figini$^{(2)}$, and Daniel C. Alexander$^{(2)}$} %\thanks{Thanks to XYZ agency for funding.}}
\address{$^{(1)}$Department of Computer Science, University of Exeter, UK\\$^{(2)}$UCL Hawkes Institute, Department of Computer Science, University College London, UK}
\def\bfDelta {{\boldsymbol{\Delta}}}

\def\bfSigma{{\boldsymbol{\Sigma}}}
\def\bfDelta{{\boldsymbol{\Delta}}}
\def\bfmu{{\boldsymbol{\mu}}}

\def\bfDelta{{\boldsymbol{\Delta}}}

\def\bs0{{\boldsymbol{0}}}
\def\bfb{{\mathbf{b}}}
\def\bfc{{\mathbf{c}}}
\def\bfd{{\mathbf{d}}}
\def\bfe{{\mathbf{e}}}

\def\bfh{{\mathbf{h}}}

\def\bfu{{\mathbf{u}}}

\def\bfw{{\mathbf{w}}}
\def\bfx{{\mathbf{x}}}
\def\bfy{{\mathbf{y}}}

\def\bfF{{\mathbf{F}}}

\def\bfI{{\mathbf{I}}}

\def\bfS{{\mathbf{S}}}

\def\bbC{{\mathbb{C}}}

\def\bbR{{\mathbb{R}}}

\newcommand{\Vpixels}{\mathbf{y}}

\newcommand{\Ndistr}[1]{\mathcal{N}_{\mathbb{R}^+}\left(#1\right)}

\newcounter{algo}
\renewcommand{\thealgo}{\arabic{algo}}

\newcommand{\bSigma}{\boldsymbol{\Sigma}}

\newcommand{\bGamma}{\boldsymbol{\Gamma}}

\begin{document}
%\ninept

\maketitle

\begin{abstract}
\vspace{-0.2cm}
We propose a novel framework for joint magnetic resonance image reconstruction and uncertainty quantification using under-sampled k-space measurements. The problem is formulated as a Bayesian linear inverse problem, where prior distributions are assigned to the unknown model parameters. Specifically, we assume the target image is sparse in its spatial gradient and impose a total variation prior model. A Markov chain Monte Carlo (MCMC) method, based on a split-and-augmented Gibbs sampler, is then used to sample from the resulting joint posterior distribution of the unknown parameters. Experiments conducted using single- and multi-coil datasets demonstrate the superior performance of the proposed framework over optimisation-based compressed sensing algorithms. Additionally, our framework effectively quantifies uncertainty, showing strong correlation with error maps computed from reconstructed and ground-truth images.
\end{abstract}

\vspace{-0.2cm}
\section{Introduction}
\vspace{-0.3cm}
Magnetic resonance imaging (MRI) is a powerful imaging modality for diagnosis of various diseases due to its high spatial resolution. Magnetic resonance (MR) image reconstruction involves solving the inverse problem using k-space data. K-space under-sampling reduces the acquisition time, leading to an ill-posed inverse problem. To address this challenge, to enable fast imaging while solving the reconstruction problem, parallel imaging \cite{beauferris2022multi, safari2025advancing} and sparse sampling \cite{hong2024complex, ahmad2020plug} are two primary approaches. These techniques help improve the reconstruction process, however, with certain limitations and challenges. One of the significant limitations is the lack of uncertainty quantification due to k-space under-sampling, which is crucial for clinical decision-making and further applications.

%The former uses multiple coils in the equipment \cite{beauferris2022multi, safari2025advancing}, while the latter breaks the Nyquist sampling barrier for sparse images \cite{hong2024complex, ahmad2020plug}. 

Bayesian inference offers the opportunity to explore the complete shape of the full joint posterior distribution using sampling methods. For example, the MCMC methods \cite{eldaly2024Bayesian, eldaly2021bayesian, jalal2021robust, levac2022accelerated} can be employed to draw samples asymptotically distributed according to the posterior distribution of interest, and therefore quantifying the uncertainty in the reconstructed images. Although Bayesian approaches for MR image reconstruction have been explored, they often involve significant simplifications for tractability. For example, \cite{bilgic2011multi, chaabene2020bayesian} transitions from enforcing sparsity in specific bases, such as wavelets and total variation, within an optimisation framework, to enforcing sparsity directly in the image domain within a Bayesian framework for tractability. In contrast, our work retains the flexibility to preserve sparsity in any given basis within a Bayesian framework without such approximations. Additionally, the Bayesian frameworks in \cite{serra2017parameter, haldar2019oedipus} primarily aim at improving sampling quality without addressing the joint image reconstruction and quantification of uncertainty using sparse priors, which is the primary focus of this work. On the other hand, while \cite{uribe2022hybrid} approximates the sparsity prior model and utilise a Metropolis-Hastings within a Gibbs sampler to sample from the resulting joint posterior distribution, these algorithms do not scale well to higher-dimensional problems like the one we consider here. Consequently, the main contributions of our work can be summarised as follows. (1) We formulate the MR image reconstruction from under-sampled k-space data within a Bayesian framework and assign a sparsity prior model to the transformed image field. No previous work attempted to sample from such joint posterior distribution using a split-and-augmented Gibbs sampler. (2) The proposed approach provides uncertainty measures corresponding to the estimated image from under-sampled k-space data, and therefore quantifying the effect of reduced samples. (3) We show that uncertainty maps can serve as a reliable tool in image interpretation by correlating them with the error maps computed between ground truth and reconstructed images. (4) The proposed algorithm allows the automated estimation of the crucial regularisation parameter which affects the resulting reconstructed image and its uncertainty bounds.

\vspace{-0.3cm}
\section{Problem Formulation}
\label{sec:ProbForm}
\vspace{-0.3cm}
The problem of MR image reconstruction can be formulated as follows: Given a set of k-space measurements of $L$ coils $\bfy \in \bbC^{NL}$, the image we aim at recovering $\bfx \in \bbR^N$ can be expressed as
\vspace{-0.1cm}
\begin{equation}
\begin{aligned}
\bfy = \bfS\bfF\mathbf{\Phi}\bfx + \bfw,
\label{eq:model}
\end{aligned}    
\end{equation}
where $\bfS$ the k-space sampling operator, $\bfF$ is the Fourier transform, $\mathbf{\Phi}$ is the coil sensitivities matrix, and $\bfw$ is the additive noise. The problem investigated in this work is to estimate the image $\bfx$ from the observation $\bfy$. This inverse problem is generally ill-posed and some sort of prior information is necessary to promote solutions with desired properties.
\vspace{-0.4cm}
\section{Hierarchical Bayesian Model}
\label{sec:bayesian}
\vspace{-0.2cm}
\noindent\textbf{Likelihood:} The likelihood function of $\bfy$ can be expressed as
\begin{equation}
\begin{aligned}
f(\bfy|\bfx) = \left(\frac{1}{2\pi \sigma^2}\right)^{\frac{NL}{2}} \exp\left(-\frac{\norm{\bfy - \bfS\bfF\mathbf{\Phi}\bfx}^2_2}{2\sigma^2}\right).
\vspace{-0.3cm}
\label{eq:modelG}
\end{aligned}
\end{equation}

The noise is assumed to be independent and identically distributed (i.i.d.) additive Gaussian noise with covariance matrix $\bSigma_\bfy = \sigma^2\bfI$, denoted as $\bfw \sim \mathcal{CN}(\bfw; \bf0 , \sigma^2\bfI)$, where $\bfI$ is the identity matrix.

\noindent\textbf{Parameter Prior Distributions:} The likelihood can limit the number of priors adequate to result in a tractable joint posterior distribution that is easy to sample from. However, with split-and-augmented (SPA) Gibbs sampler \cite{pereyra2023split, vono2019split}, a variety of priors can be used as long as there exists a proximal operator to them. In this work, we consider sparsity in the spatial gradient of the reconstructed image using a total varation (TV) prior model, which can be written as $p(\bfx|\tau) = \frac{1}{\theta} \exp\left(-\tau \text{TV}(\bfx)\right),$ where $\theta = \int \tau \text{TV}(\bfx) d\bfx$, and $\tau$ is the regularisation parameter. In this model, the regularisation parameter $\tau$ is also supposed to be unknown and is included in the inference process.% using the stochastic approximation proximal gradient algorithm (SAPG) \cite{vidal2020maximum}.

\noindent\textbf{Joint Posterior Distribution:} The joint posterior of the parameter vector $\bfx$ and hyperparameter $\tau$ can be expressed as 
$$\hspace{-4.5cm}p(\bfx, \tau|\Vpixels) \propto p(\Vpixels|\bfx)p(\bfx|\tau)$$
\begin{equation}
\begin{aligned}
\vspace{-0.7cm}
\propto \exp\left(-\frac{\norm{\bfy - \bfS\bfF\mathbf{\Phi}\bfx}^2_2}{2\sigma^2}\right)\times \exp\left(-\tau \text{TV}(\bfx)\right).
%p(\bfx, \tau|\Vpixels)&\propto p(\Vpixels|\bfx)p(\bfx|\tau)\propto \left(\frac{1}{2 \pi \sigma^2}\right)^{\frac{NL}{2}} \exp\left( - \frac{\| \mathbf{y} - \mathbf{S} \mathbf{F} \MATproj \mathbf{x} \|_2^2 }{2 \sigma^2} \right) \\ & \times \exp\left(-\tau \text{TV}(\bfx)\right).
\label{eq:posteriornew}
\end{aligned}
\vspace{-0.4cm}
\end{equation}
\section{Bayesian Inference}
\label{sec:BayesianInference}
\vspace{-0.2cm}
To overcome the challenging derivation of Bayesian estimators associated with $f(\bfx,\tau|\bfy)$, we use an efficient MCMC method to generate samples asymptotically distributed according to Eq. \eqref{eq:posteriornew}. More precisely, we consider a variable-splitting inspired MCMC algorithm, namely the split-and-augmented Gibbs sampler (SPA) that was recently proposed in \cite{pereyra2023split, vono2019split}. For SPA, splitting variables are introduced to decouple the likelihood from the prior distribution, i.e. $\bfb =\bfx$, $\bfc = \bfS\bfd$, $\bfd = \bfF\bfe$, and $\bfe = \mathbf{\Phi}\bfx$. This splitting decouples locally each vector of variables from the rest of the variables, which will in turn make the inference easier. Thus, the joint posterior distribution in Eq. \eqref{eq:posteriornew} can be extended as 
\newcommand\numberthis{\addtocounter{equation}{1}\tag{\theequation}}
\vspace{-0.2cm}
$$p(\bfx, \bfDelta, \bGamma| \bfy) \propto \exp\left(-\frac{\norm{\bfy - \bfc}^2_2}{2\sigma^2}\right) \times \exp\left(-\tau \text{TV}(\bfb)\right)$$
\vspace{-0.2cm}
$$\times \exp\left(-\frac{\norm{\bfb - (\bfx+\bfh_1)}_2^2}{2\rho^2}\right) \times \exp\left(-\frac{\norm{\bfc - (\bfS\bfd+\bfh_2)}_2^2}{2\rho^2}\right)$$
\vspace{-0.2cm}
$$\times \exp\left(-\frac{\norm{\bfd - (\bfF\bfe+\bfh_3)}_2^2}{2\rho^2}\right)  \times \exp\left(-\frac{\norm{\bfe - (\mathbf{\Phi}\bfx+\bfh_4)}_2^2}{2\rho^2}\right)$$
\vspace{-0.4cm}
\begin{align*}
\times \exp\left(-\frac{\left(\norm{\bfh_1}_2^2+\norm{\bfh_2}_2^2+\norm{\bfh_3}_2^2+\norm{\bfh_4}_2^2\right)}{2\alpha^2}\right), \numberthis \label{eq:PosteriorSPA}
\vspace{-0.5cm}
\end{align*}
where $\rho, \alpha > 0$, and $\bGamma = \{\bfh_1$, $\bfh_2, \bfh_3, \bfh_4\}$ are auxiliary variables associated with $\bfDelta = \{\bfb$, $\bfc, \bfd, \bfe\}$. We can observe that it's easy to sample from the conditional distribution of each of the parameters. We propose to sample sequentially $\bfx$, $\bGamma$, $\bfDelta$, and $\tau$ using moves that are summarised in Algorithm \ref{algo:PCGS}, with the details of sampling of each variable presented below.
\vspace{-0.5cm}

%The stochastic simulation method presented in this work provides information of the full joint posterior distribution. However, in the provided experiments, we focus on two posterior statistics, the marginal posterior mean of each pixel, which is the minimum mean squared error estimator (MMSE), and the marginal posterior variance \cite{robert2007bayesian}. 

\begin{algorithm}
\caption{Split-and-Augmented Gibbs Sampling for Joint MR Image Reconstruction and Uncertainty Quantification}
\label{algo:PCGS}
\begin{algorithmic}[1]
\State \textbf{Fixed input parameters}: Number of burn-in iterations $N_{\text{bi}}$, total number of iterations $N_{\text{MC}}$
\State \textbf{Initialisation} ($k = 0$)
\begin{itemize}
\item Set $\bfb^{(0)}, \bfc^{(0)}, \bfd^{(0)}, \bfe^{(0)}, \bfh_1^{(0)}, \bfh_2^{(0)}, \bfh_3^{(0)}, \bfh_4^{(0)}, \tau^{(0)}$
\end{itemize}
\State \textbf{Repeat ($1 \leq k \leq N_{\text{bi}}$)}
\item Sample $\tau^{(k)} | {\bfx}^{(k)}$ % using Eq. \eqref{eq:updateTau}
%\State \textbf{Set} $k = k + 1$.
\State \textbf{Repeat ($N_{\text{bi}}+1 \leq k \leq N_{\text{MC}}$)}
\item Sample $\bfx^{(k)} | \left(\bfy, \bfDelta^{(k-1)}, \bGamma^{(k-1)}, {\tau}^{(k-1)}\right)$% using Eq. \eqref{eq:SampleX}
\State Sample $\bfb^{(k)} | \left(\bfx^{(k)}, \bfc^{(k-1)}, \bfd^{(k-1)}, \bfe^{(k-1)}, \bGamma^{(k-1)}\right)$ %using Eq. \eqref{eq:SampleB}
\State Sample $\bfc^{(k)} | \left(\bfy, \bfx^{(k)}, \bfb^{(k)}, \bfd^{(k-1)}, \bfe^{(k-1)}, \bGamma^{(k-1)}\right)$ %using Eq. \eqref{eq:SampleC}
\State Sample $\bfd^{(k)} | \left(\bfy, \bfx^{(k)}, \bfb^{(k)}, \bfc^{(k)}, \bfe^{(k-1)}, \bGamma^{(k-1)}\right)$ %using Eq. \eqref{eq:SampleC}
\State Sample $\bfe^{(k)} | \left(\bfy, \bfDelta^{(k)}_{\backslash \bfe}, \bGamma^{(k-1)}\right)$% using Eq. \eqref{eq:SampleD}
\State \textbf{Repeat ($1 \leq i \leq 4$)}
\State Sample $\bfh_i^{(k)} | \bfDelta^{(k)}$% using Eq. \eqref{eq:SampleH1}
%\State \textbf{Set} $k = k + 1$.
\end{algorithmic}
\end{algorithm}
\vspace{-0.3cm}

Sampling $\bfx | \left(\bfy, \bfDelta, \bGamma, {\tau}\right)$ reduces to sampling from a multivariate Gaussian distribution 
$\mathcal{N}\left({\bfx; \bfmu_\bfx, \bfSigma_\bfx}\right),$ with $\bfmu_\bfx = \frac{1}{\rho^2}\left(\mathbf{\Phi}^T(\bfe - \bfh_4) + \left(\bfb-\bfh_1\right)\right)\bfSigma_\bfx$ and $\bfSigma_\bfx = \rho^2\left(\mathbf{\Phi}^T\mathbf{\Phi} + I \right)^{-1}$. Similarly, sampling each of $\bfc | \left(\bfy, \bfx, \bfDelta_{\backslash \bfc}, \bGamma\right)$, $\bfd | \left(\bfy, \bfx, \bfDelta_{\backslash \bfd}, \bGamma\right)$, and $\bfe | \left(\bfy, \bfx, \bfDelta_{\backslash \bfe}, \bGamma\right)$ reduces to sampling from a complex multivariate Gaussian distribution with $\bfmu_\bfc = \frac{\rho^2\bfy + \sigma^2(\bfS\bfd+\bfh_2)}{\sigma^2 + \rho^2}$, $\bfmu_\bfd = \frac{1}{\rho^2}\left(\bfS^T(\bfc-\bfh_2) + \bfF\bfe+\bfh_3\right)\bfSigma_\bfd$, $\bfmu_\bfe = \rho^2(\mathbf{\Phi}\bfx + \bfh_4$ $+ \bfF^C(\bfd-\bfh_3))\bfSigma_\bfe$, $\bfSigma_\bfc = \frac{\rho^2\sigma^2}{\rho^2+\sigma^2}\bfI$, $\bfSigma_\bfd = \rho^2\left(\bfS^T\bfS + \bfI_N\right)^{-1}$, and $\bfSigma_\bfe = \rho^2\left(\bfF^C\bfF + \bfI\right)$. On the other hand, for sampling $\bfb | \left(\bfy, \bfx, \bfDelta_{\backslash \bfb}, \bGamma\right)$, it can be seen from Eq. \eqref{eq:PosteriorSPA} that the full conditional distribution of $\bfb$ reduces to: $\exp\left(-\tau \text{TV}(\bfb)\right) \times \exp\left(-\frac{\norm{\bfb - (\bfx+\bfh_1)}_2^2}{2\rho^2}\right).$ Due to the non-differentiability of this conditional distribution, we consider the proximal Moreau-Yoshida-unadjusted Langevin algorithm (P-MYULA) \cite{durmus2018efficient} to generate samples asymptotically distributed according to this distribution. The Markov chain of P-MYULA is written as
\begin{equation}
\begin{aligned}
\bfb^{(k+1)} &= (1 - \frac{\gamma}{\lambda})\bfb^{(k)} - \gamma \nabla f(\bfb^{(k)}) + \frac{\gamma}{\lambda}\text{prox}_g^\lambda(\bfb^{(k)})\\ &+ \sqrt{2\gamma}\,\,\Ndistr{\bf0, \bf1},
\label{eq:SampleB}
\end{aligned}
\end{equation}
where $f(\bfb) = \frac{\norm{\bfb - (\bfx+\bfh_1)}_2^2}{2\rho^2}$, that is $\nabla f(\bfb) = \frac{\bfb - \bfx -\bfh_1}{\rho^2}$
and $g(\bfb) = \tau \text{TV}(\bfb)$, that is $\text{prox}_g^\lambda(\bfb) = \arg \min_\bfu \tau\lambda \text{TV}(\bfu) + \frac{1}{2}\norm{\bfu - \bfb}_2^2$, which can be solved itratively using Chambolle's algorithm \cite{chambolle2004algorithm}. Sampling each of {$\bfh_1 | \left(\bfx, \bfDelta\right)$}, {$\bfh_2 | \bfDelta$}, {$\bfh_3 | \bfDelta$}, and \textit{{$\bfh_4 | \left(\bfx, \bfDelta\right)$}} reduces to sampling from a multivariate Gaussian distribution $ { \mathcal{N}\left({\bfh_i; \bfmu_{\bfh_i}, \bfSigma_{\bfh_i}}\right),}$ with $i = \{1, 2, 3, 4\}, $ and $\bfmu_{\bfh_1} = \frac{\alpha^2}{\rho^2+\alpha^2}\left(\bfx - \bfb\right)$, \text{    } $\bfmu_{\bfh_2} = \frac{\alpha^2}{\rho^2+\alpha^2}\left( \bfS\bfd - \bfc\right)$, $\bfmu_{\bfh_3} = \frac{\alpha^2}{\rho^2+\alpha^2}\left( \bfF\bfe - \bfd\right)$, $\bfmu_{\bfh_4} = \frac{\alpha^2}{\rho^2+\alpha^2}\left( \mathbf{\Phi}\bfx - \bfe\right)$, and $\bfSigma_{\bfh_1}=\bfSigma_{\bfh_2}=\bfSigma_{\bfh_3} = \bfSigma_{\bfh_4}=\frac{\rho^2\alpha^2}{\rho^2+\alpha^2}\bfI$. Finally, sampling the regularisation parameter $\tau|\bfx$ is done using the stochastic approximation proximal gradient algorithm (SAPG) proposed in \cite{vidal2020maximum}. This method computes the maximum marginal likelihood using a stochastic proximal gradient optimisation algorithm that is driven by proximal MCMC samplers. The update of $\tau | \bfx$ is given as $\tau^{(t+1)} = \Pi_T\left[\tau^{(t)} + \delta^{(t+1)} \left( \frac{NL}{\tau^{(t)}} - \text{TV}(\bfx^{(t)})\right)\right],$ where $\Pi_T$ is the projection onto $T$ that is $\tau\in T$ and $\delta$ is a sequence of non-increasing step-sizes set as in \cite{vidal2020maximum, bobkov2011concentration}.% In this work, we set $\delta = \frac{0.1}{NL\times\tau^{(0)}}\times t^{-0.8}$, and $T \in [10^{-5}, 1]$ as in \cite{vidal2020maximum, bobkov2011concentration}.

The algorithm is stopped after $N_{\textrm{MC}}$ iterations, including $N_{\textrm{bi}}$ burn-in iterations which correspond to the transient period of the sampler. The first $N_{\textrm{bi}}$ samples are discarded and the remaining samples are used to approximate the empirical average of the generated samples, e.g., the MMSE of the actual intensity vector $\bfx$ given by $\hat{\bfx} = \frac{1}{N_{\text{MC}} - N_{\text{bi}}}\sum_{t = N_{\text{bi} + 1}}^{N_{\text{MC}}} \bfx^{(t)}.$ and the marginal variance (quantified uncertainty).
\vspace{-0.4cm}
\section{Experiments and Results}
\label{sec:Simulations}
\vspace{-0.3cm}
In order to assess the performance of the proposed approach for joint MR image reconstruction and uncertainty quantification using sub-sampled k-space measurements, we consider single- and multi-coil datasets. The single coil dataset comprises 100 T1-weighted brain test images from the human connectome project (HCP), in which ground truth is available \cite{sotiropoulos2013advances}. For multi-coil data, we use 10 T2-weighted k-space test images from the M4Raw real low-field MRI k-space data set which consists of multi-channel brain k-space data acquired on a 0.3T MRI system \cite{lyu2023m4raw}. The size of images in both datasets is $256\times 256$. The data acquisition is simulated by sub-sampling the k-space of the two datasets, using 2D-random sampling as in \cite{korkmaz2022unsupervised, zhang2020image, qu2024radial}. Cartesian and pseudo-radial sampling provide similar behaviour and therefore the results are not included here. In our framework, the under-sampling ratio is set as \{5\%, 10\%, 20\%, 30\%, 40\%\}. Our reconstruction algorithm, which we call MCMC-TV, is compared against three methods: (1) IFFT, the baseline inverse Fourier transform; (2) ADMM-TV, an optimisation-based counterpart of our approach that maximises the log-joint posterior distribution at fixed hyperparameters using the alternating direction method of multipliers (ADMM); and (3) ADMM-Wav, another optimisation-based method that enforces sparsity in the wavelet domain. These optimisation-based methods (ADMM-TV and ADMM-Wav) are widely used in the standard compressed sensing community and are also based on a variable splitting strategy \cite{fessler2020optimization}. The quantitative measure used to assess the quality of reconstructed images is the root mean square error (RMSE). For the MCMC-TV method, Markov chains of length $N_{\text{bi}} = 1.7\times 10^4$ and burn-in periods of length {$N_{\text{MC}} = 2\times 10^4$} are used. Different values of the hyperparameters $(\rho, \alpha)$ of the SPA-Gibbs sampler are tested to maximise the performance, and then fixed for all experiments. The hyperparameters of the P-MYULA proximal MCMC method are set to $(\lambda, \gamma) = (\rho^2, \rho^2/4)$ as in \cite{vono2019split}.
\vspace{-0.4cm}
\subsection{Results using single-coil data}
\vspace{-0.25cm}
Table \ref{tab:HCPMetrics} provides RMSE measures of the estimates of the four methods at increasing under-sampling ratio using the HCP test images. In general, IFFT produces the poorest reconstruction results, as indicated by the highest RMSE values, and the MCMC-TV method consistently delivers the best reconstruction results, with the lowest RMSE values for all k-space ratios, except for the very high under-sampling rate (5\% sampling), where the counterpart optimisation-based method ADMM-TV provides better results. In terms of reconstruction quality using the different sampling ratios, we can observe that as the under-sampling ratio decreases, the reconstruction quality reduces for all of the tested methods.

\begin{table}%[]
\centering
\small
\setlength{\tabcolsep}{1pt}
\caption{Average and standard deviation of RMSE using different under-sampling ratios. Bold indicates best, and underlined indicates second best results.}
\label{tab:HCPMetrics}
\begin{tabular}{|c|ccccc|}
\hline
\multirow{2}{*}{} &
  \multicolumn{5}{c|}{\textbf{Under-sampling Ratio}} \\ \cline{2-6} 
\textbf{Method} &
  \multicolumn{1}{c|}{\textbf{5\%}} &
  \multicolumn{1}{c|}{\textbf{10\%}} &
  \multicolumn{1}{c|}{\textbf{20\%}} &
  \multicolumn{1}{c|}{\textbf{30\%}} &
  \multicolumn{1}{c|}{\textbf{40\%}} \\ \hline \hline
\textbf{IFFT} & 
\multicolumn{1}{c|}{19.85$\pm$2.8} & 
\multicolumn{1}{c|}{15.32$\pm$2.5} & 
\multicolumn{1}{c|}{12.08$\pm$2.2} & 
\multicolumn{1}{c|}{9.72$\pm$1.8} & 
\multicolumn{1}{c|}{7.67$\pm$1.5} \\ \hline 
\textbf{ADMM-Wav} & 
\multicolumn{1}{c|}{5.62$\pm$1.5} & 
\multicolumn{1}{c|}{3.71$\pm$1.2} & 
\multicolumn{1}{c|}{2.26$\pm$1.0} & 
\multicolumn{1}{c|}{1.45$\pm$0.8} & 
\multicolumn{1}{c|}{\underline{0.98$\pm$0.6}} \\ \hline 
\textbf{ADMM-TV} & 
\multicolumn{1}{c|}{\textbf{5.56$\pm$1.5}} & 
\multicolumn{1}{c|}{\underline{3.68$\pm$1.2}} & 
\multicolumn{1}{c|}{\underline{2.25$\pm$1.0}} & 
\multicolumn{1}{c|}{\underline{1.42$\pm$0.8}} & 
\multicolumn{1}{c|}{\underline{0.98$\pm$0.6}} \\ \hline 
\textbf{MCMC-TV} & 
\multicolumn{1}{c|}{\underline{5.65$\pm$1.6}} & 
\multicolumn{1}{c|}{\textbf{3.60$\pm$1.2}} & 
\multicolumn{1}{c|}{\textbf{2.11$\pm$1.0}} & 
\multicolumn{1}{c|}{\textbf{1.30$\pm$0.8}} & 
\multicolumn{1}{c|}{\textbf{0.90$\pm$0.6}} \\ \hline  
\end{tabular}
\vspace{-0.5cm}
\end{table}
\begin{table}%[h]
\small
\setlength{\tabcolsep}{3pt}
\caption{Average correlation coefficient (CC) computed between marginal standard deviation and error maps.}
%\vspace{-0.2cm}
\label{tab:CorrHCP}
\centering
\begin{tabular}{|c|c|c|c|c|c|}
\cline{1-6}
\textbf{SR} & \textbf{5\%} & \textbf{10\%} & \textbf{20\%} & \textbf{30\%} & \textbf{40\%} \\ \hline\hline
\multicolumn{1}{|l|}{\textbf{\begin{tabular}[c]{@{}l@{}}CC\end{tabular}}}
& \begin{tabular}[c]{@{}c@{}}0.80$\pm$.068\end{tabular} & 
\begin{tabular}[c]{@{}c@{}}0.79$\pm$.059\end{tabular} & 
\begin{tabular}[c]{@{}c@{}}0.79$\pm$.051\end{tabular} & 
\begin{tabular}[c]{@{}c@{}}0.75$\pm$.047\end{tabular} & 
\begin{tabular}[c]{@{}c@{}}0.74$\pm$.045\end{tabular} \\ \hline
\end{tabular}
\vspace{-0.5cm}
\end{table}

\begin{figure}%[h]
    \centering
    \vspace{-0.2cm}
    \includegraphics[width=0.43\textwidth]{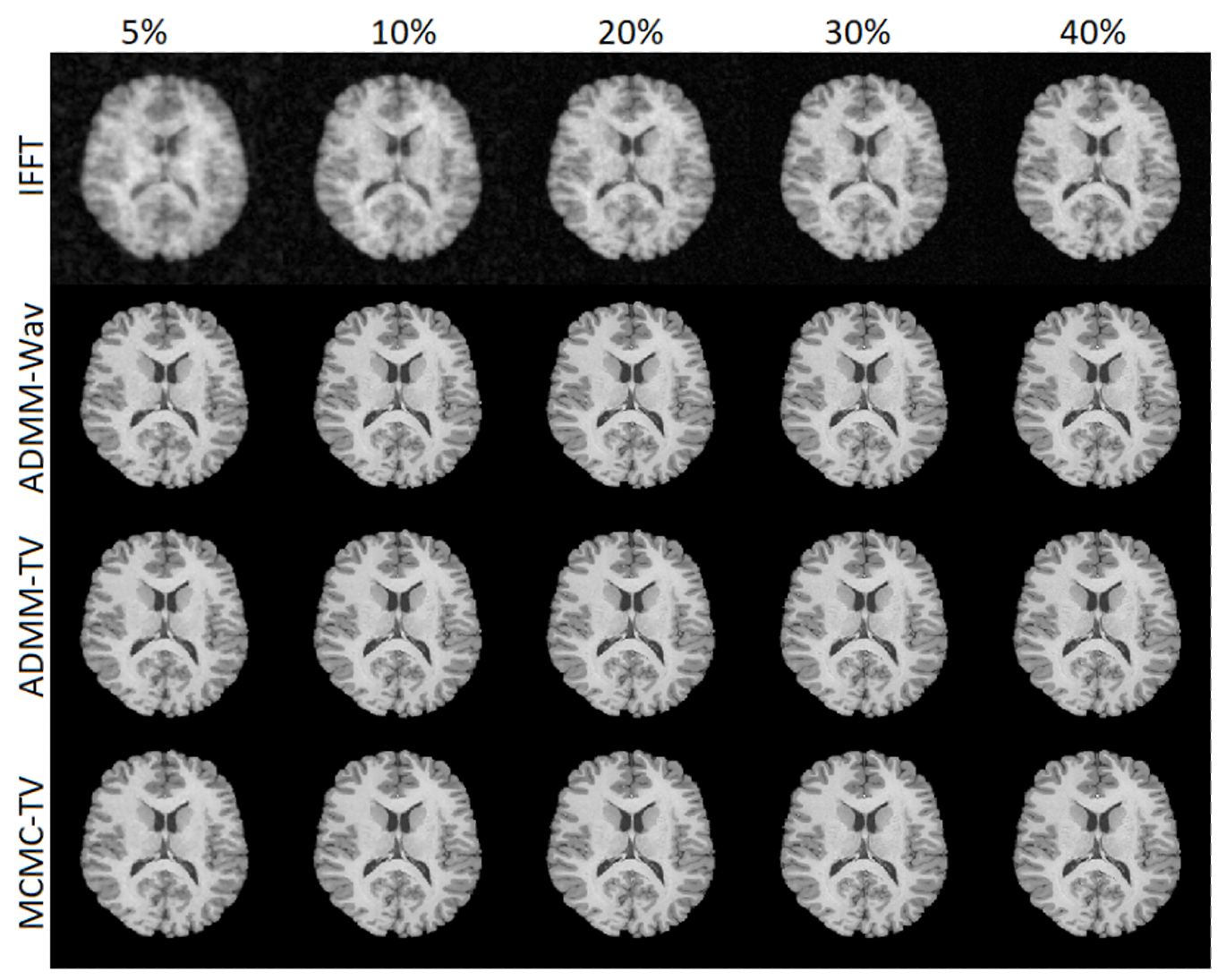}
    \vspace{-0.5cm}
    \caption{Results of (a) image reconstruction of a brain test image from HCP data set using the four tested methods at increasing under-sampling ratio.}
    \label{fig:MMSEEstimates}
\end{figure}
%\vspace{-0.2cm}
\begin{figure}%[h]
    \centering
    \vspace{-0.2cm}
    \includegraphics[width=0.43\textwidth]{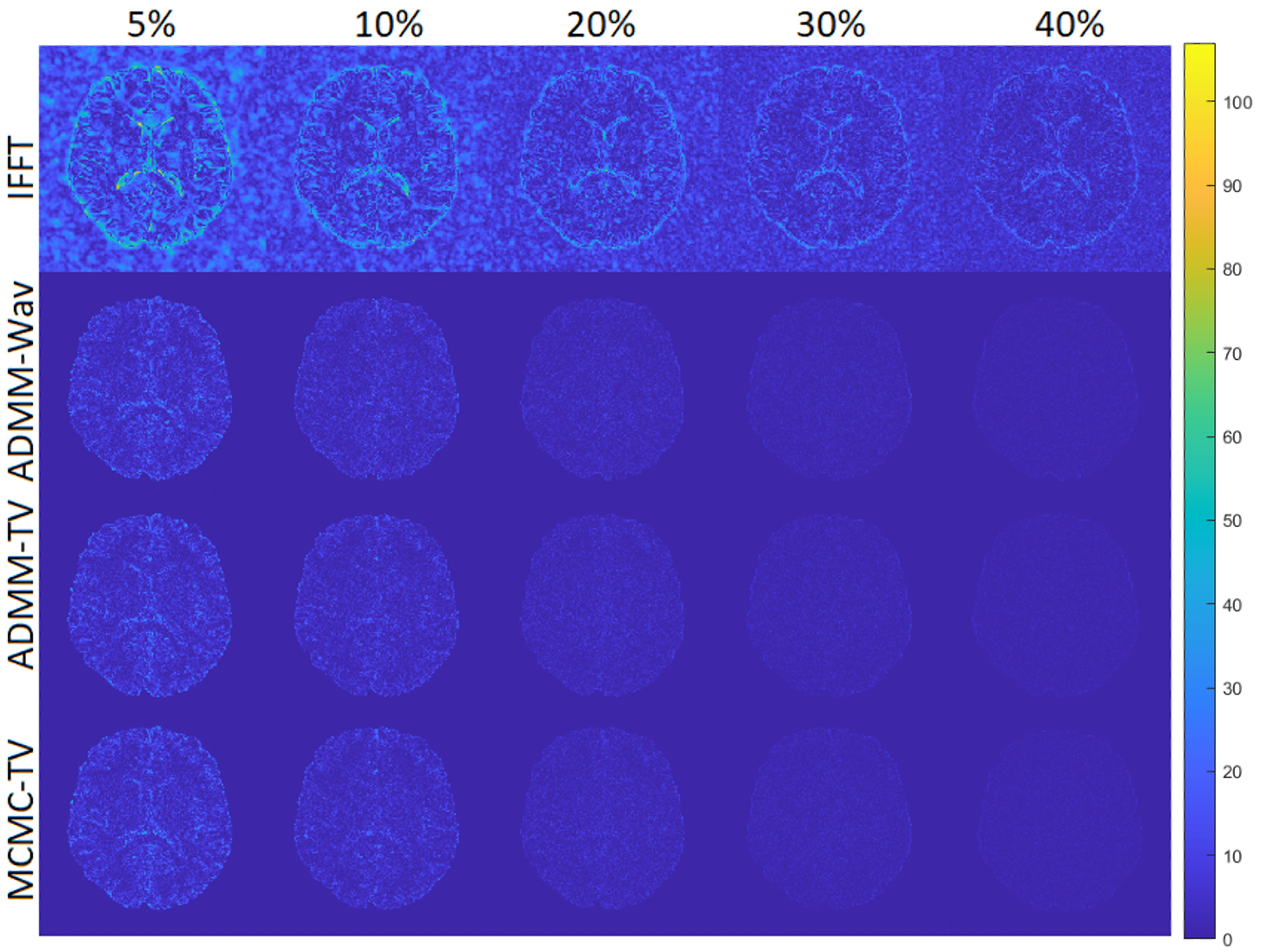}
    \vspace{-0.5cm}
    \caption{Error map images computed between ground truth and image estimates in Fig. \ref{fig:MMSEEstimates}.}
    \label{fig:Errors}
\end{figure}
%\vspace{-0.1cm}
\begin{figure}%[h]
    \centering
        \vspace{-0.3cm}
    \includegraphics[width=0.43\textwidth]{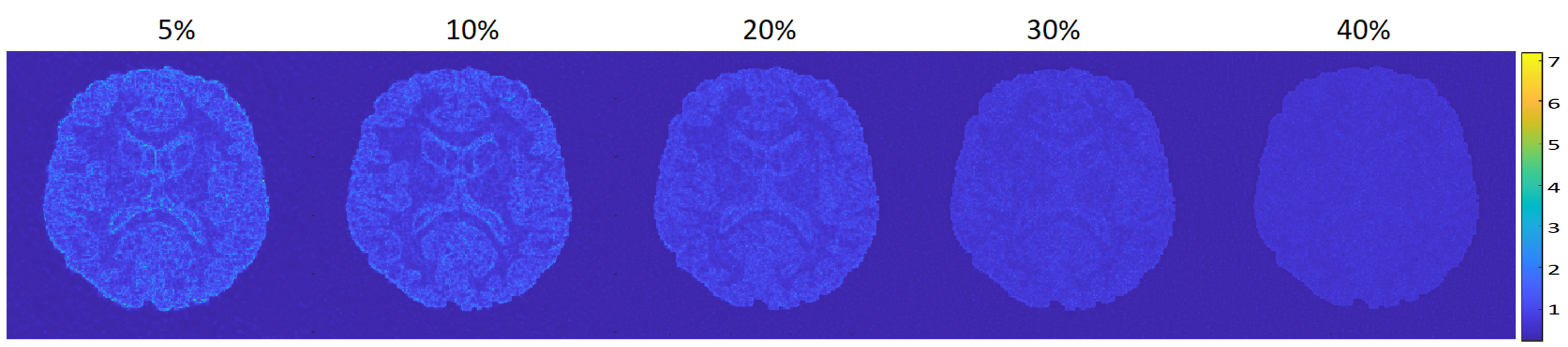}
        \vspace{-0.5cm}
    \caption{Marginal standard deviation of the MCMC-TV estimates in Fig. \ref{fig:MMSEEstimates}.}% Rest of the methods provide point-based estimates and therefore don't quantify the uncertainties of the reconstructed images.}
    \label{fig:UQ}
\end{figure}
%\vspace{-0.2cm}
\begin{figure}%[h]
    \centering
    \includegraphics[width=0.43\textwidth]{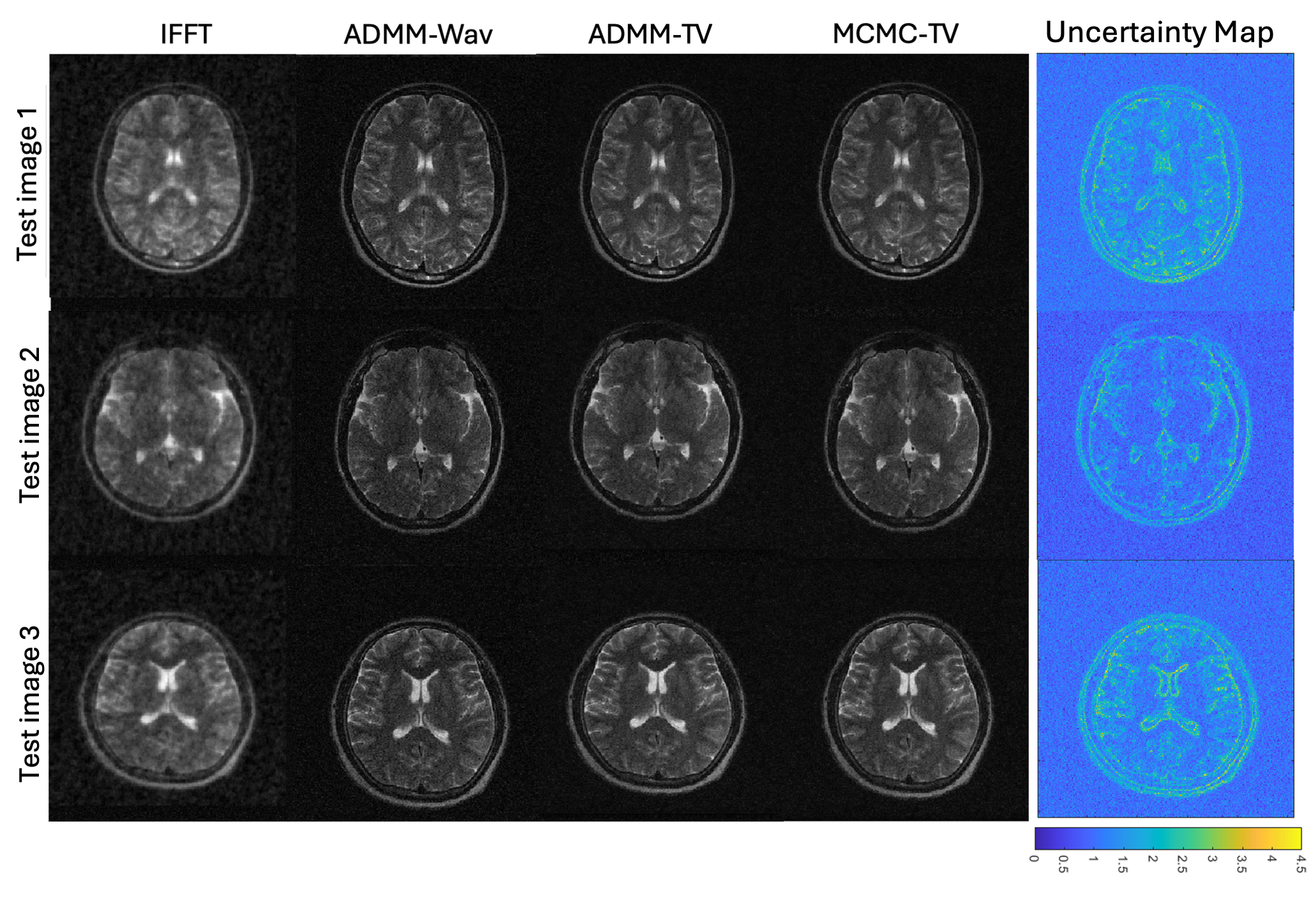}
    \vspace{-0.5cm}
    \caption{Results of reconstruction of three brain test images from the real low-field MRI M4Raw using multicoil data at 20\% under-sampling ratio.}
    %\vspace{-0.4cm}
    \label{fig:Miccai3}
\end{figure}

Figure \ref{fig:MMSEEstimates} shows the reconstruction results of an example test image and the error map images between the original clean image and the reconstructed ones, using the four approaches at increasing under-sampling ratio. The regularised methods produce reconstructions with significantly fewer artifacts compared to IFFT. Among these, the MCMC-TV approach delivers the most satisfactory results (except at very high sampling ratio), characterised by clear contours, sharp edges, and fine image details. This is particularly evident in the error map images, where the structural information is more coherent, especially near the major artifacts present in the IFFT reconstructions. These qualitative observations align with the earlier quantitative results.

The proposed MCMC-TV method draws samples asymptotically distributed according to the joint posterior distribution, and therefore these samples can be used to quantify the bounds of their posterior mean. Figure \ref{fig:UQ} shows the marginal standard deviation of the MCMC-TV method at increasing under-sampling ratio. The results show that as the sampling ratio increases, the uncertainty of the estimates also increases, and vice versa. The uncertainty maps correlate well with the error maps shown in Figure \ref{fig:MMSEEstimates} for the MCMC-TV method. As the k-space under-sampling ratio increases, the errors become more pronounced, and the uncertainty of the estimates rises correspondingly. This relationship is quantified in Table \ref{tab:CorrHCP}, which shows the correlation coefficients between the error map images and the marginal standard deviations. We can observe higher correlation between the error maps and the marginal standard deviation. This finding aligns with the quantitative results discussed earlier, where MCMC-TV demonstrated superior quantitative performance. The notable correlation with the calculated errors supports the hypothesis that the estimated uncertainty effectively captures the true uncertainty, making it a reliable metric for aiding image interpretation in practice when no ground truth is available.

\vspace{-0.4cm}
\subsection{Results using multi-coil data}
\vspace{-0.3cm}
We applied the proposed approaches and the three baseline methods to reconstruct real low-field MRI k-space data from the M4Raw data set. Due to similar performance as to the previous experiments, we report results for the 20\% sub-sampling rate. Since no ground truth is available, we present image reconstruction results for all methods, alongside the corresponding uncertainty maps for the MCMC-TV method. Figure \ref{fig:Miccai3} displays the reconstructed images of the four tested methods as well as the uncertainty maps of MCMC-TV reconstruction. We can observe that while the optimisation-based methods yield competitive reconstructions to the MCMC-TV method, they only provide point estimates and lack uncertainty quantification in contrast to the Bayesian method MCMC-TV which offers uncertainty bounds. The results indicate higher uncertainty near intensity transitions, consistent with expected behavior, while regions of constant intensity exhibit lower uncertainty.

\vspace{-0.3cm}
\section{Conclusion}
\label{sec:Conclusion}
\vspace{-0.3cm}
In this work, a joint image reconstruction and uncertainty quantification framework from under-sampled k-space MRI is proposed. The problem is formulated within a Bayesian framework, and the reconstructed image is assigned a total-variation prior model. Bayesian inference was performed using a split and augmented-Gibbs sampler, while the non-smooth terms were sampled using a proximal MMCMC method. The proposed model outperformed the counterpart optimisation-based methods, and also provided uncertainty measures to the estimates which cannot be obtained using the optimisation methods. Thus, the main focus of this work was to provide credibility intervals to the estimates and correlating them with error maps between reconstructed and ground truth images, unlike most methods in the literature which focus more on the absolute image reconstruction quality. Future work may consider combining the two prior models, TV and wavelets for better reconstruction results.

\bibliographystyle{IEEEbib}
\bibliography{strings,refs}

\end{document}